\documentclass[manuscript]{aastex}

\usepackage{xspace}
\usepackage{graphicx}

\renewcommand\arcdeg{\mbox{$^\circ$}\xspace}%
\renewcommand\micron{\mbox{$\mu$m}\xspace}%

\newcommand{\kms}{${\rm km\;s}^{-1}$\xspace}
\newcommand{\vlsr}{V_{lsr}}
\newcommand{\tpm}{$\pm$}

\newcommand{\ms}{$M_\sun$\xspace}

\newcommand{\mdot}{$\dot{M}$\xspace}
\newcommand{\rave}{\langle r \rangle}
\newcommand{\msunyr}{$M_\sun$~yr$^{-1}$\xspace}

\newcommand{\sect}[1]{\S\ref{#1}\xspace}

\renewcommand{\micron}{$\mu$m\xspace}
\newcommand{\um}{\micron}
\newcommand{\isotope}[2]{$^{#1}$C$^{#2}$O\xspace}

\slugcomment{High Velocity Outflows in V Hydrae}
\shorttitle{}
\shortauthors{Sugerman, Sahai \& Hinkle}

\begin{document}

\title{Sculpting an AGB Mass-Loss Envelope into a Bipolar Planetary Nebula: High-Velocity
Outflows in V Hydrae}

\author{ Raghvendra
Sahai\altaffilmark{1}, Ben E.K.~Sugerman\altaffilmark{2}, Kenneth
Hinkle\altaffilmark{3}}
\altaffiltext{1}{Jet Propulsion Laboratory, California Institute of Technology, 4800 Oak Grove
Dr., Pasadena,
CA 91109}
\altaffiltext{2}{Goucher College, 1021 Dulaney Valley Rd
Baltimore, MD 21204}
\altaffiltext{3}{National Optical Astronomy Observatory, PO Box 26732, Tucson, AZ 85726}

\authoremail{raghvendra.sahai@jpl.nasa.gov}

\begin{abstract}
We have carried out high-resolution spectroscopic observations of the carbon star V Hya,
covering the 4.6\um band of CO. These data, taken over 7 epochs, show that the circumstellar
environment of V Hya consists of a complex high-velocity 
(HV) outflow containing at least six kinematic components with
expansion velocities ranging between 70 and 120 \kms, together with  
a slow-moving normal outflow at about 10\,\kms. Physical changes occur in the HV outflow
regions
on a time-scale as short as two days, limiting their extent to be $\lesssim10^{16}$
cm. The intrinsic line-width for each HV component is quite large
($6-8$\,\kms) compared to the typical values ($\sim1$
\kms) appropriate for normal AGB circumstellar envelopes (CSEs), due to excess turbulence
and/or large velocity gradients resulting from the energetic interaction of the HV outflow
with the V Hya CSE. We have modelled the absorption features to set constraints on the
temperature distribution in, and the mass ejection-rates for gas in the main HV components.

\end{abstract}

\keywords{stars: AGB and post-AGB, stars: mass-loss, stars: carbon, ISM: jets and outflows,
stars: individual (V Hydrae)}

\section{Introduction}
Stars of mass about ($1-8$)\,\ms~live uneventful lives on the main sequence. But as AGB stars,
they
eject, over a period of 10$^4$-10$^5$ years or less, half or more of their mass in slow
spherical
winds. And then, in a very poorly understood short (few 100--1000 yr) phase, they are
transformed first into aspherical pre-planetary nebulae (PPNs), which then evolve into
aspherical planetary nebulae (PNs). Recent morphological studies with the Hubble
Space Telescope ($HST$) support the idea
that high-speed, jet-like outflows play a crucial role in this transformation (Sahai \&
Trauger 1998, Sahai et al. 2007, Balick \& Frank 2002). These outflows may be driven by an
accretion disk around a binary companion (Morris 1987). However, evidence for such outflows is
indirect; this phase is so short that few nearby stars are likely to be caught in the act.

One such star is carbon-rich V Hydrae (V Hya), well known for its collimated, high-velocity
($\gtrsim$50 \kms) outflows. The unusual nature of the mass ejections in V Hya was first
noticed in single-dish observations of the CO J=1-0 line: e.g., \citep{KMJ88} concluded that
the V Hya circumstellar envelope (CSE) consists of a primary outflow which is roughly
isotropic, and expanding slowly (emitting over a velocity range of about 16 \kms), as well as
a secondary bipolar outflow, which is expanding roughly twice as fast. Very high-velocity
outflows were first seen via observations of individual P and R-branch lines in the
fundamental vibration-rotation band of \isotope{12}{16} (hereafter CO) at 4.6 \um. These data,
taken over two epochs by \citet[][hereafter SW88]{SW88} showed the presence of several
absorption components due to high-velocity (hereafter HV) gas up to $V_{exp} \approx 120$
\kms. The absorption lines showed significant differences between the two epochs, implying
changes in the physical properties of the HV outflow over year-long time-scales. This
study provided the first direct evidence of the simultaneous presence of a slowly expanding
envelope and such a high-velocity stellar wind in a mass-losing AGB carbon star. \citet{Eva91}
found outflow velocities up to 155 \kms from optical spectroscopy, while \citet{LB92} found
3.6 cm continuum emission suggesting the presence of shock-heated gas in the HV outflow. High
signal-to-noise millimeter-wave CO J=2--1 and 3--2 line profiles showed wide wings, indicating
the presence outflowing gas at velocities of $\gtrsim$200\,\kms~(Knapp et al. 1997: KJY97).
Interferometric mapping of the
CO J=2--1 \& 3--2 emission resolved the high-velocity emission and showed that it came from an
extended,
bipolar and highly collimated outflow (Hirano et al. 2004: Hetal04). The
optical spectrum of V Hya shows anomalously broad photospheric lines, which Barnbaum, Morris
\& Kahane (1995) interpret in terms of rotation due to spin-up by a companion in a common
envelope configuration, although alternative explanations have been presented (Luttermoser \&
Brown 1992, KJY97). 

Sahai et al. (2003a: Setal03) used the Space Telescope Imaging Spectrograph onboard the $HST$
to image highly excited HV material: they found a blob of hot, shocked gas
(emitting in the [SII]$\lambda$4069.7\AA~line) moving away from the central star with a
projected radial velocity of -240\,\kms~relative to the systemic velocity. Setal03 also found a
compact ($\approx0.5{''}$), hot, slowly-expanding (10--15\,\kms) central disk-like structure,
as well as a large ($\approx20{''}$), cool, equatorially-flattened structure. Although the
expansive
kinematics of the former implies that it is not an accretion disk, Setal03 speculate that it
may result from a recent phase of equatorially enhanced mass-loss, which may be enhancing the
accretion process. In a succesful search for binary companions of cool AGB stars via the 
far-UV emission from the former, Sahai et al. (2008) find that V Hya stands out amongst their
survey objects as having the largest FUV flux, as well as the highest FUV-to-NUV flux ratio.
They conclude that the FUV emission, which is many orders of magnitude larger than the
contribution from the carbon star, is either directly or indirectly due to a companion star --
photospheric emission from the companion in the first case, or an accretion disk around the
companion, in the second case.

V\,Hya is thus the best example to date of an AGB star with an active, collimated outflow, 
dense equatorially-flattened structures possibly related to a central accretion disk, and an
inferred binary companion.

In this paper, we report high-resolution, high signal-to-noise spectroscopy of the 4.6 \um band
of CO and its isotopomers in V Hya, taken over 7 epochs spanning 2 years, in order to
characterize and constrain the kinematic structure and time-variations of the HV outflow, and
its hydrodynamic interaction with the normal envelope. Observations and data reduction are
described in \sect{sec-obs}, and the observational results are presented in
\sect{sec-results}. We interpret the spectra in terms of a multi-component high-velocity
outflow in \sect{sec-interp}. Deconvolution of the kinematic structure of this outflow and
detailed modeling of its components is presented in \sect{sec-analy}. We examine and discuss
discussing the implications of our results for the final stages of the transformation of an
AGB star into a PN in \sect{sec-disc}. We present our conclusions in \sect{sec-concl}.

%Must add info about mm-wave spectra, velocity structure in the mm CO
%lines, how that is not easily deconvolved into low and high-velocity
%outflows.  
%Add Knapp et al 1997 and Hirano et al 2004.  

\section{Observations}\label{sec-obs}

\begin{figure}
\hskip -1.0cm
%\rotatebox{270}{\resizebox{0.7\textwidth}{!}{\includegraphics{cotile}}}
\rotatebox{270}{\resizebox{0.71\textwidth}{!}{\includegraphics{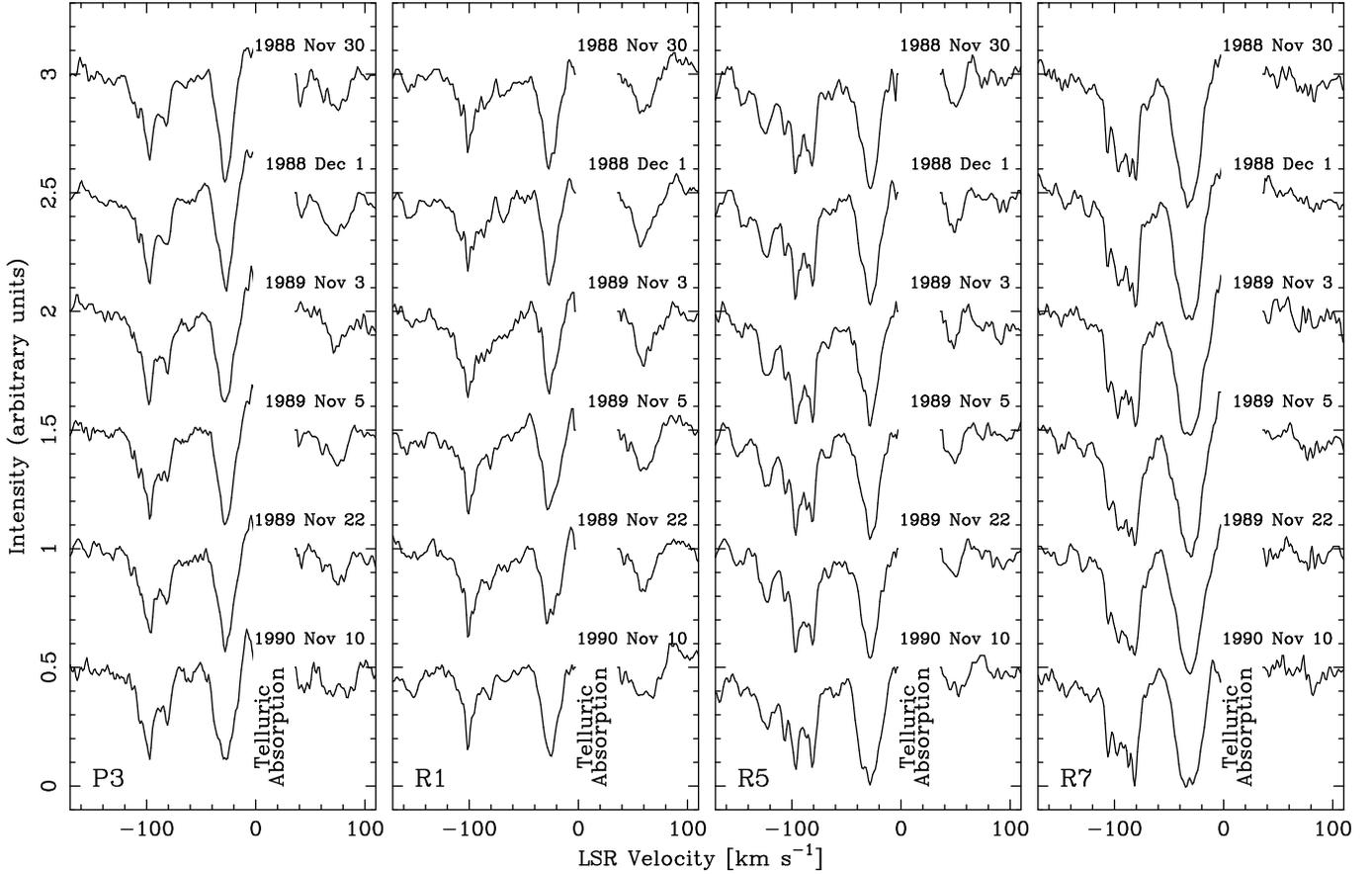}}}
\caption{Sample spectra of different CO $(v=1-0)$
  lines (P3, R1, R5, R7) and different 
  epochs. Regions of strong telluric
  absorption have been blanked out. The continuum
  levels for the spectra have been scaled to unity, but constant
offsets have then been added to them, in order
to shift them vertically for clarity in this plot. These offsets are (from bottom to top) -0.5,
0, 0.5, 1.0, 1.5, \& 2.0.
\label{cotile}}
\end{figure}

Using the Fourier Transform Spectrometer at the coud\'e focus of the Kitt Peak 4-m Mayall
telescope, we observed the 4.6 \um lines of the fundamental CO vibration-rotation band at 7
different epochs (1988 Nov 30, 1988 Dec 1, 1989 Jan 19, 1989 Nov 3, 1989 Nov 5,
1989 Nov 22, and 1990 Nov 10) in the 2100--2200 cm$^{-1}$ range, at an unapodized resolution
of 0.02 cm$^{-1}$; the spectra were oversampled and recorded at a resolution of 0.0089
cm$^{-1}$.
Data were reduced as described in \citep{SW85}. Briefly, each spectrum
was divided by the instrumental frequency-response determined by observing a tungsten lamp
source. The spectra were then corrected for telluric absorption at the air mass of the source
observation as derived from a lunar spectrum divided by the lamp source and an empirically
derived 314\,K blackbody function for the Moon. We have used $\vlsr=-16$ \kms for the stellar
radial velocity, $V^*$, a value which lies between earlier measurements ($-14$ to $-16$\,\kms:
Knapp \& Morris 1985, Zuckerman \& Dyck 1986, KJY97) and more recent ones ($-17.5$\,\kms:
Kahane et al. 1996, Hetal04). The spectra have been normalized to a local
continuum level of unity with an uncertainty level of $\sim$0.05. Representative CO ($v=0-1$)
spectra (P3, R1, R5, and R7) from each epoch (except 1989 Jan 19, for which the Moon spectrum
for telluric correction was not available) are shown in Figure \ref{cotile}.

\section{Results}\label{sec-results}

\subsection{CO Absorption Features}\label{sec-results-selec}

\begin{figure}
\rotatebox{270}{\resizebox{0.7\textwidth}{!}{\includegraphics{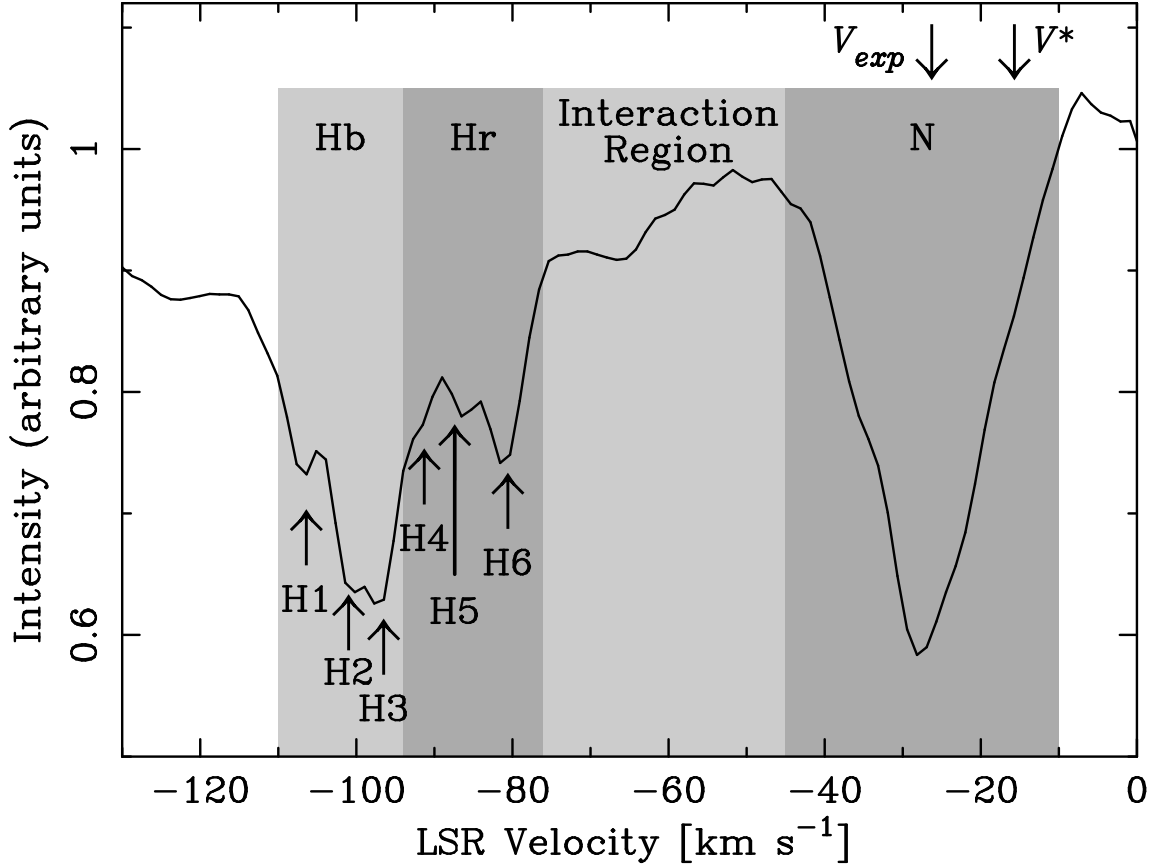}}}
 \caption{Sample CO spectrum produced by averaging the R1, R2, and R5
 spectra over all epochs, showing the various velocity regions and individual high-velocity
components H1--H6. The continuum
 level for each spectra was scaled to unity before averaging. 
 Shading denotes
 broader velocity regions discussed in the text (e.g.\,\sect{sec-results}).  The LSR velocities
of V Hya and the peak
 absorption of the normal component (``N") are marked by $V^*$ and $V_{exp}$,
 respectively.  
\label{hlabel}}
\end{figure}

We first examine the qualitative absorption-line characteristics of the most common CO
isotopomers. The signatures (absorption features) of the different outflows (e.g., the normal
and HV outflows) are cleanly separated in the 4.6\,\um~spectra (Fig. \ref{hlabel}), unlike the
case of mm-wave CO emission spectra taken with telescopes whose beams do not adequately
resolve the emitting region. The CO ($v=0-1$) lines show a component extending from roughly
$\vlsr\approx-15$ to $-43$ \kms with peak fractional absorption of roughly 50--60\%. This is
the normal component (hereafter labeled ``N'') resulting from absorption in the
slowly-expanding CSE around V Hya. The peak absorption in the normal component occurs around
$\vlsr\sim -28$ \kms, implying an expansion velocity for the slow outflow near 12 \kms.
However, the normal component shows definite structure -- shoulders can be seen in each wing
of the CO ($v=0-1$) normal component -- indicating the presence of kinematic perturbations
superimposed on the slow spherical outflow. Of note, the blue wing of the normal component is
anomalously wide compared to typical values (few tenths to 1 \kms) believed to be appropriate
for the CSEs of mass-losing AGB stars (see, e.g., Sahai 1990, Teyssier et al. 2006). The
half-power point of the normal absorption feature lies roughly 9 \kms beyond the velocity of
peak absorption, significantly larger as the values observed in other AGB stars (less than a
few\,\kms: Sahai 1990).

\begin{figure}
\rotatebox{270}{\resizebox{0.9\textwidth}{!}{\includegraphics{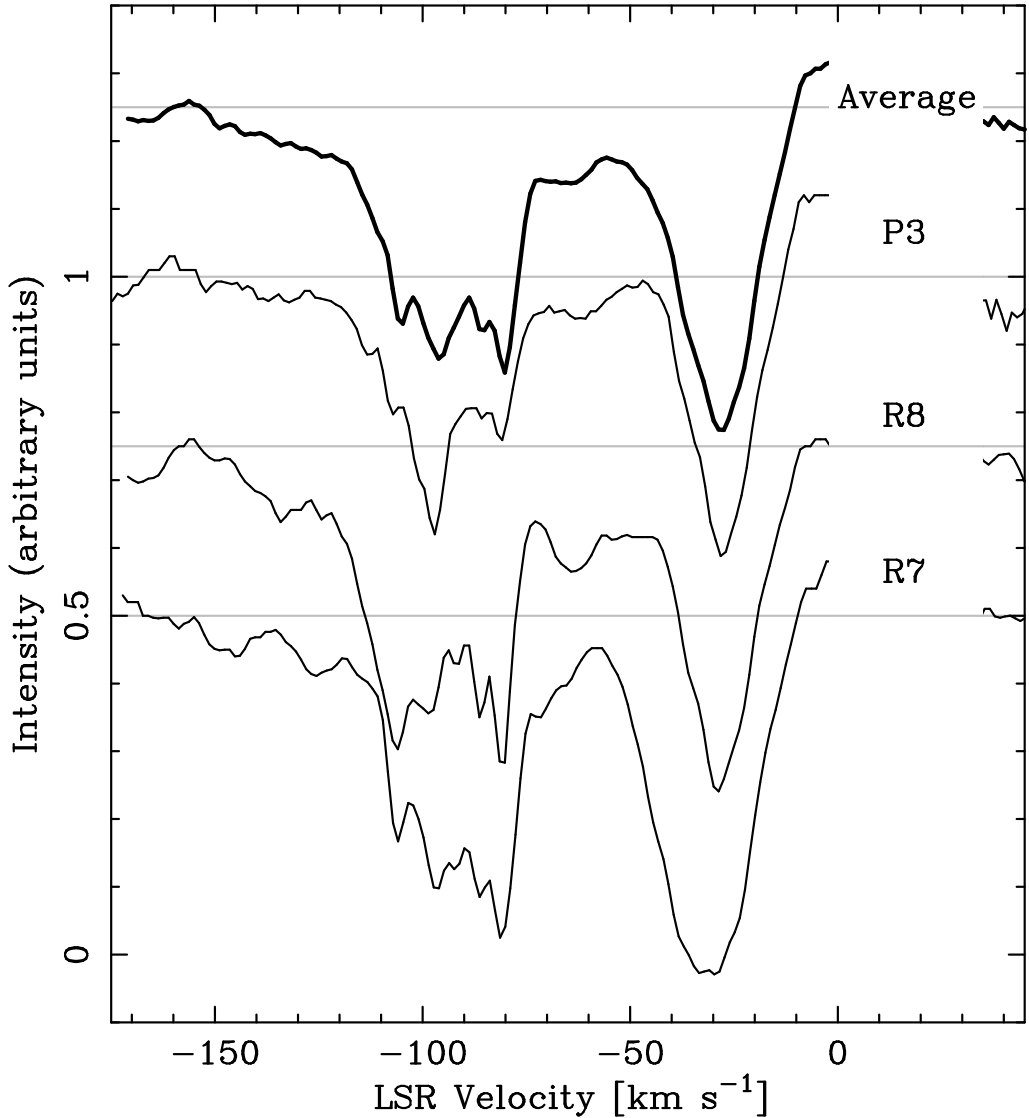}}}
 \caption{CO P3, R7, and R8 spectra averaged over all epochs, plus the
 average of these spectra, showing the highest-velocity features
 extending blueward from H1 ($V_{lsr}\approx -110$ \kms) to about -155
 \kms. The continuum levels for the spectra have been scaled to unity, but constant offsets
have then been added, in order
to shift them vertically for clarity in this plot. These offsets are -0.5 (R7), -0.25 (R8), 0
(P3), \& 0.25 (Average). 
\label{fastvel}}
\end{figure}

In addition to the normal component, we see prominent absorption in the $\vlsr\approx -75$ to
$-115$ \kms region, with a fractional absorption of 5--50\%. This absorption arises in gas
expanding at much larger velocities than in the slow outflow. The HV absorption feature is
complex, with at least six discrete, partially overlapping features, which imply the presence
of a corresponding number of distinct kinematic components. These have been labeled H1 through
H6, with H1 the bluemost feature (Fig. \ref{hlabel}). There is also weaker absorption extending
blueward of $\vlsr
\approx -115$ \kms, which is best seen in the spectra of the P3, R8, and R7 lines
(Fig.\,\ref{fastvel}). Each of these have been chosen as they have no contaminating
lines\footnote{i.e., overlapping lines from the $v=0-1$ band of other CO isopotomers, such as
$^{13}$CO, C$^{17}$O, or C$^{18}$O, or the $v=1-2, 2-3$,... bands of $^{12}$CO and its
isopotomers} bluewards of the H1 component, up to $\vlsr=-175$ \kms. The HV absorption
``tail'' extending to about $\vlsr=-155$ \kms is clearly seen in the high signal-to-noise
average of the three spectra. Thus we conclude that the HV outflow contains gas expanding at
velocities up to 140 \kms. Weak absorption can also be discerned in the spectral region
between the normal and HV outflows ($\vlsr=-45$ to $-70$ \kms).

\begin{figure}
\rotatebox{270}{\resizebox{1.0\textwidth}{!}{\includegraphics{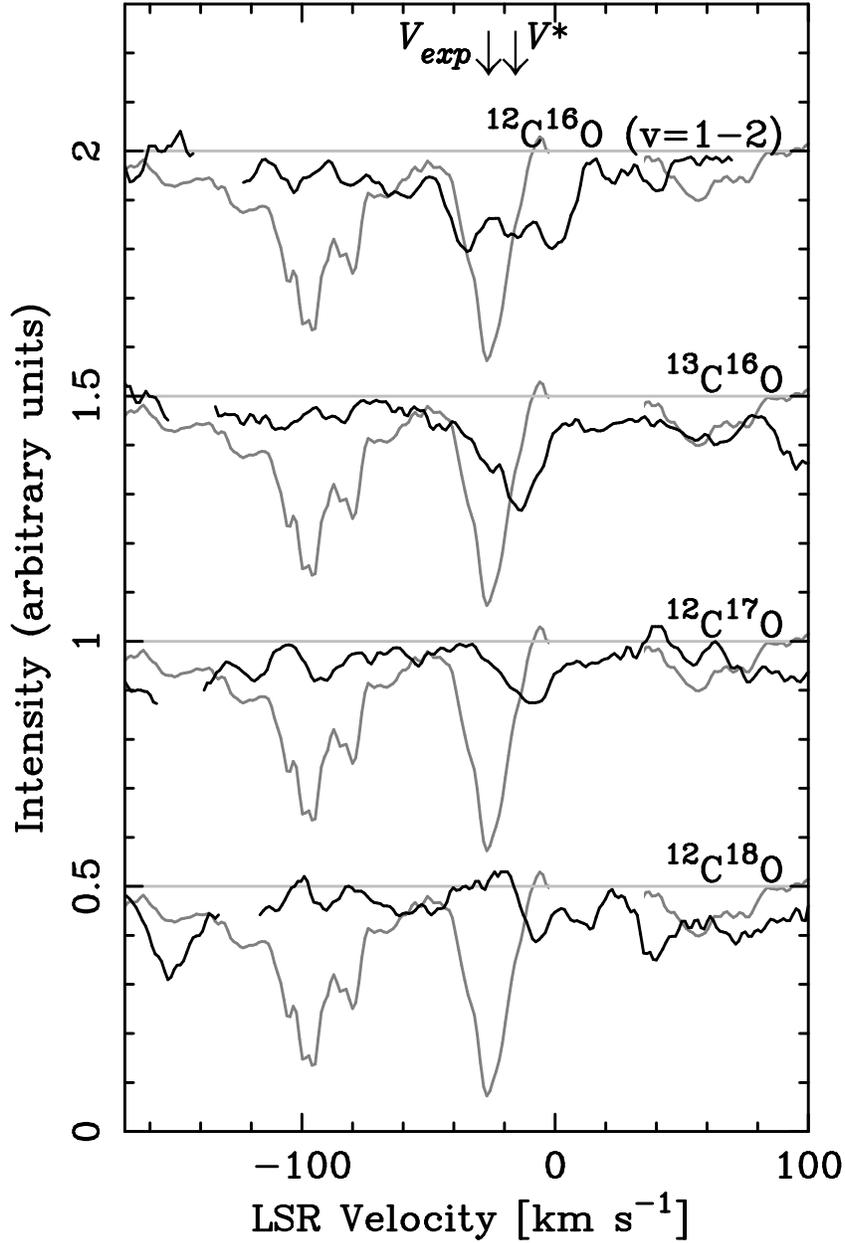}}}
 \caption{Average spectra showing the absorption features of CO
 $(v=1-2)$, \isotope{13}{}, \isotope{}{17}, and \isotope{}{18},
 as marked in each panel, plotted over the average CO $(v=1-0)$
 spectrum (in grey, from Fig.\,\ref{hlabel}). Regions of strong telluric
 absorption have been blanked out.  LSR velocities of V Hya and the peak
 absorption of the normal component are marked (as in Fig.\,\ref{hlabel}). The continuum levels
(in light grey)  
for the spectra have been scaled to unity, but constant offsets have then been
added, in order to shift them
vertically for clarity in this plot. These offsets are (from bottom to top) -0.5, 0, 0.5, \&
1.0.
\label{iso}}
\end{figure}

In order to be able to select the least contaminated CO lines for analysis, we examine the
absorption strengths of $v=0-1$ lines from other isotopomers, as well as the $v=1-2$ lines of
CO. In Figure \ref{iso}, we show averages of 2 to 4 individual lines for each of the possible
contaminants. The lines for each contaminant have been selected to include only those which do
not have overlapping features from CO or other isopotomers in the normal or HV regions. The
strongest two contaminants are CO $(v=1-2)$ and \isotope{13}{} $(v=0-1)$. 

Within the velocity interval covered by the CO ($v=0-1$) normal component, the CO $(v=1-2)$
line shows two local absorption minima, at LSR velocities of $-35$ \kms~and $-14$ \kms. These
absorption features occur at the same velocities as the shoulders in the wings of the CO
($v=0-1$) normal component, but are significantly stronger than the absorption due to the
normal component of CO $(v=1-2)$, and thus provide us a method for measuring their center
velocities accurately. This is important, because even though these components are
only present weakly in the wings of the CO ($v=0-1$) normal component, they can potentially
shift the apparent velocity of the peak of the cool normal component from its actual value. We
have therefore first made Gaussians fits to the CO $(v=1-2)$ normal absorption features to
determine the precise velocity centers of the $-35$ and $-14$ \kms absorption features, and
using these centers as fixed inputs, we have made a 3-gaussian fit to the CO ($v=0-1$) normal
absorption feature. This yields a peak absorption velocity of $V_{lsr}=-26.3$ \kms for the
main feature, giving an outflow velocity of 10\,\kms for the slow outflow.

\isotope{13}{} ($v=0-1$) has a normal absorption component with
maximum absorption depth of 20\%, and only marginal HV absorption.
\isotope{}{17} ($v=0-1$) has weak normal absorption with a fractional
depth of 10\% and no obvious HV features. There are no detected
features for the \isotope{}{18} molecule.

Given these restrictions, we chose spectral windows with
$\Delta\nu=2.6$ cm$^{-1}$ ($\Delta V=360$ \kms) around individual
absorption lines which are free from strong telluric lines or
absorption features from contaminants.  We have chosen the CO ($v=1-0$) R1, R2,
R5, R7, R8, P1 and P3 lines.  However, there are small regions of the spectral windows for these
lines which are contaminated.  The normal components of CO ($v=1-0$) P1
and P3 are contaminated by atmospheric absorption and by CO ($v=1-2$)
R3, respectively.  The normal feature of CO ($v=1-0$) R7 is overlapped by CO
($v=1-2$) R15.  The normal component of CO ($v=1-2$) R13 partially
overlaps the H1 component of CO ($v=1-0$) R5, causing it to broaden.  The HV
region of CO ($v=1-0$) R8 is unusable due to overlapping CO ($v=1-2$) R16 and R24
absorption.  Therefore, we have not used the normal absorption
features of CO ($v=1-0$) R7, P1, and P3 in the following analyses.

\subsection{Time Variations} \label{sec-results-variation}

\begin{figure}
\rotatebox{270}{\resizebox{1.0\textwidth}{!}{\includegraphics{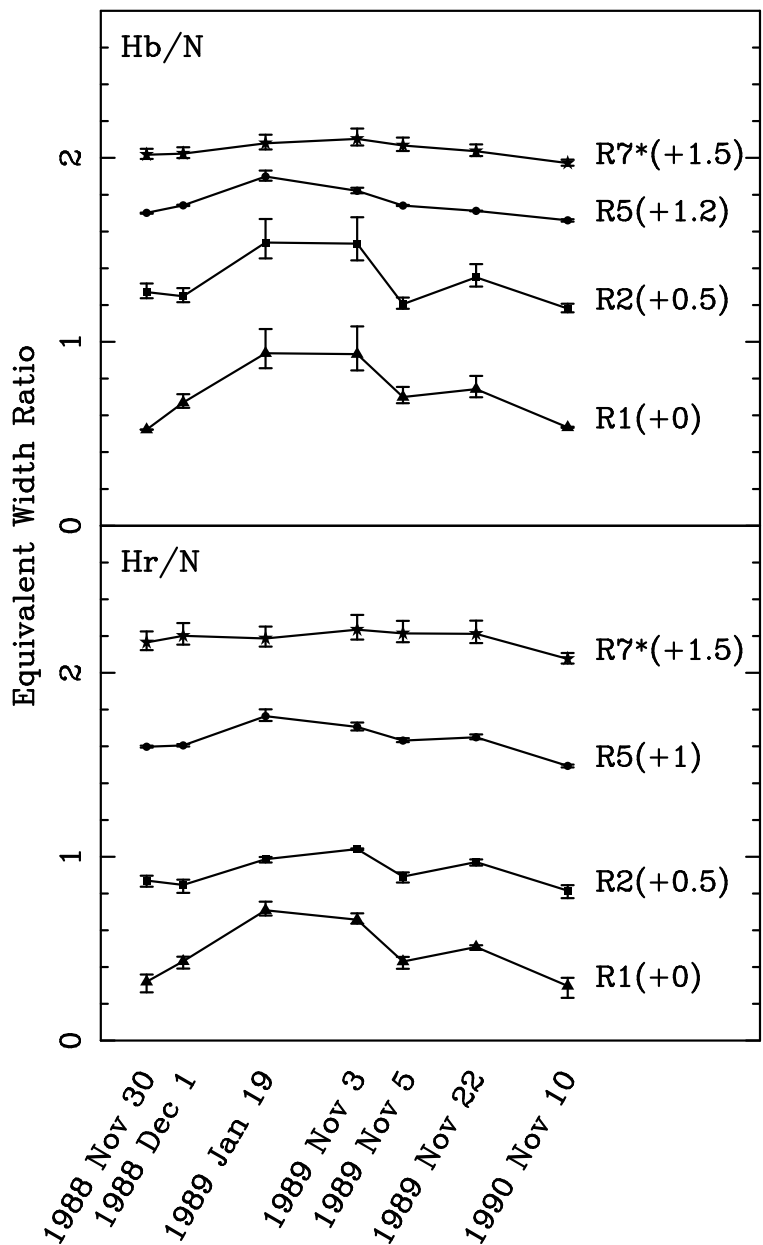}}}
 \caption{Equivalent width ratios of the high-velocity blue (Hb) and
 red (Hr) components (as defined in the text and Fig.\ \ref{hlabel})
 to that of the normal component. Constant offsets have then been
added to each of these  (as shown) in this plot, in order to shift them
vertically for clarity. Epochs are denoted at bottom, and
 the rotational transitions at right.  
  \label{vhgr}}
\end{figure}

The high-velocity absorption features were found to change with time in the SW88
study, over the 2 epochs (separated by slightly over 1 year) at which spectra were
taken. In order to search for time variations in the outflows in our data, we have
calculated the ratios of
absorption intensities, integrated over selected velocity bins, in the CO ($v=1-0$)
R1, R2, R5 and R7*\footnote{the * denotes our use of the normal absorption
component from the R8 line as a proxy for the same in R7, due to the overlap of the CO
($v=1-2$) R15 line in the normal component region of R7} lines for each epoch of
observation\footnote{epoch 1989 Jan 19 has been included here because the ratios of the
equivalent widths are not affected by the lack of Moon spectra for telluric correction for
this epoch}.
We define three velocity bins, labeled N, Hb, and Hr
(see Fig.\,\ref{hlabel}). The ``N'' bin spans the normal component
from $\vlsr=-10$ to $-40$ \kms.  Hb and Hr cover the HV
region, with Hb spanning the three bluemost features H1, H2, and H3
($\vlsr=-110$ to $-94$ \kms) and Hr spanning the three remaining
redward components H4, H5, and H6 ($\vlsr=-94$ to $-76$ \kms).  Errors
in the integrated intensities have been calculated based on
uncertainty in the continuum level and RMS noise.  We show the ratios
of Hr and Hb to N for each epoch and transition in Figure \ref{vhgr}.
Dramatic variations can be seen in the ratios from one epoch to the
other, including the shortest time scale of two days sampled between
1988 Nov 30 and Dec 1.  These variations show that significant changes
are occuring in one or both outflows over very short timescales. This timescale 
limits the size of the region where the changes are occuring to
$\lesssim10^{16}$ cm, based on the light-crossing time. Thus, the changes most likely arise in
the HV outflow, given that the N component is due to absorption in the extended mass-loss
envelope of V Hya. It is clear that the changes in the low-J lines (R1, R2) are much more
pronounced than in the high-J lines (R5, R7). We have no simple explanation of this effect. We note
that infrared pumping due to the 4.6\micron~continuum is expected to affect the population of the
low-J levels more than the high-J ones (Schoenberg 1988), so it is possible that fast variations in
the 4.6\micron~continuum may play a role in producing this effect: a detailed excitation
calculation is required to assess its importance.
Numerical simulations of an episodic, collimated high-velocity outflow interacting with an
extended AGB mass-loss envelope are needed to investigate if sufficiently rapid local changes in
density and temperature can be produced which may contribute to this effect.

\section{The Multi-Component Molecular Outflow in V Hya} \label{sec-interp}

Our near-infrared data, together with millimeter-wave and optical spectroscopy, 
show the presence of multiple molecular outflows in V Hya. As better
(high angular resolution, higher sensitivity) millimeter-wave CO data have 
become available, authors have constructed increasingly detailed models of these
outflows on the
basis of the line profiles and maps (Kahane et al. 1988, 1996, KJY97, Hetal04). The
most recent of these is by Hetal04, which is based on interferometric mapping of the
CO J=2-1 emission. This model has 3 kinematic components, defined as a  (i)
low-velocity, expanding, flattened  ``disk-like" envelope, with $\Delta
V=\pm$8\,\kms~around the systemic velocity, (ii) a medium-velocity wind, with
$\Delta V=\pm$60\,\kms, and (iii) a high-velocity jet, with $\Delta
V=\pm(60-160)$\,\kms. Hetal04 show that the axis of the high-velocity jet is
perpendicular to the major axis of the envelope and the medium-velocity wind. 
%In Figure \ref{apjtoon}, we schematically show the simplest
Since Hetal04 only recover $\sim$50\% of the single-dish CO J=2-1 flux in their maps,
it is clear that they have resolved out a more extended component, which  presumably
is the slowly expanding, spherical mass-loss CSE. KJY97, from their single-dish
spectra of CO J=3-2, CS J=5-4 and HC$_3$N (25-24) lines, identify such an extended
component, expanding at an (uncertain) speed of 15\,\kms, as the ``normal" slow AGB
wind. 

The normal absorption feature arises in the region where the line of sight intersects the
slowly-expanding normal CSE (resolved out in the SMA data: Hetal04) and the flattened
disk-like envelope. The HV features arise in the region where the line of sight crosses the HV
outflows. The absorption occuring at outflow velocities between those of the normal and HV
components could arise in the region where the HV outflows interact with the slow outflow; we
therefore call it the ``interaction region''. Although the expansion velocity represented by
this feature apparently suggests that it is associated with the medium-velocity wind described
by Hetal04, such an interpretation is inconsistent with the blue-shift of the ``interaction
region'' features. This is because the medium-velocity wind is red-shifted on the side of the
star that shows the blue-shifted HV outflow (and which is responsible for the HV features in
our data).
The strength of the absorption in the interaction region increases from CO R1 to R8, showing
that it arises in highly-excited gas. High excitation temperatures are expected in this region
because of deposition of kinetic energy of the HV outflow as it is decelerated by the slow
wind.

\subsection{High-Velocity Outflow Temperatures} \label{sec-interp-temp}

\begin{figure}
\rotatebox{270}{\resizebox{0.6\textwidth}{!}{\includegraphics{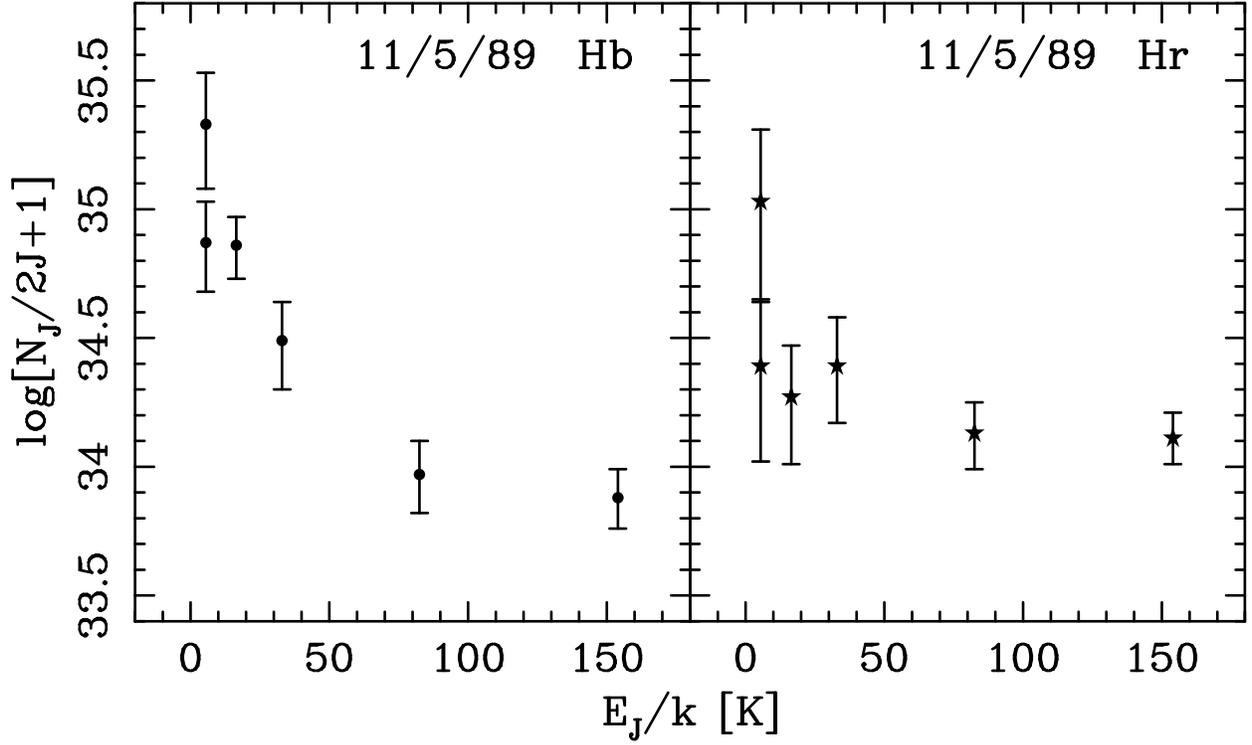}}}
 \caption{Populations of individual $J$ levels, calculated using the
equivalent width of the corresponding transitions for the Hb and Hr
components, plotted against $E_J/k$.  Note that the R1 and P1 data
points ($E_J/k=5.5$), which represent the population of the $J=1$
level and should coincide, are significantly separated.
  \label{popj}}
\end{figure}

The integrated absorption intensity over the full HV region
increases by a factor of 1.6 from CO R1 to R5, and by 1.2 from R5 to R7.
Thus the absorption arises in a region of highly-excited
gas.  Assuming that the absorption is optically thin, we can use the
linear part of the curve of growth to calculate the temperatures for
each of the three integrated intensity components.  The populations in
the individual $J$ levels have been calculated using the equivalent
width of the corresponding transitions, and plotted against $E_J/k$.  
An example of such a plot is shown in Figure \ref{popj}.
If LTE prevails, then all the points should fall on a straight line
with a slope equal to $-1/T$ where $T$ is the temperature.  Clearly
the data cannot be fitted with a straight line, implying that the
absorbing column of gas cannot be characterized by a single
temperature. Nevertheless, a straight-line fit gives us a rough estimate of the
excitation temperatures in the outflow components, yielding
$T_{Hb}=117\pm9$ K and $T_{Hr}=243\pm61$ K. Many authors have
chosen to fit such plots using two-temperature fits, with one linear
fit to the low-$J$ points, and another to the high-$J$ data. However, we adopt a more
sophisticated approach, described below.  Such an approach is necessary,
not only because a
single-temperature fit is incorrect, but more importantly, because the
R1 and P1 data points, which represent the population of the $J=1$
level and should coincide, are significantly separated, implying that
these lines are not optically thin, i.e., we cannot assume that $\tau << 1$. 

\section{Detailed Modelling} \label{sec-analy} 

We now describe a model in which the absorbing column is characterised by a
continuous density and temperature distribution. 
We assume
that each of the HV absorption features arises in a shell of gas
expanding radially away from the star.  Each shell is assumed for
computational convenience to have spherical curvature, a constant
expansion velocity, an inverse square molecular density dependence on
radius, and a power-law excitation temperature characterizing the
rotational ladder. 
We define each shell in terms of 6 parameters: (1) the mass-loss
rate \mdot, (2) the mean shell radius $\rave$, (3) the fractional shell
thickness $\delta r$ (i.e., the outer and inner  
radii of each shell are $r_\pm=\rave (1 \pm \delta r/2$), (4) the excitation
temperature $T_{\rave}$ at
radius $\rave$, (5) the power-law exponent for the temperature
variation $\beta$ such that $T(r)=T_{\rave}(r/\rave)^{-\beta}$, and (6)
the intrinsic line width $\Delta V_t$ (set by turbulent broadening). The parameters \mdot and
$\rave$ are
degenerate, since if each of these is scaled by the same factor, the total absorption
column density, $N$, remains unchanged. Hence we can only determine the ratio
\mdot/$\rave$ from our modelling. 

The calculation of the equivalent width of
each P- or R-branch line for an adopted set of model parameters is 
explained in Appendix\,\sect{app-ew}. 
For each kinematic component modelled, the
expansion velocity is fixed at the value
determined from Gaussian line-fitting, described in Appendix\,\sect{obs-ew}.  
%The distance to the source is taken to be 390 pc 
%\citep[close to 380 pc reported by][]{Kna97}
The relative abundance (by number) of CO to H$_2$, $f_{CO}$, is fixed at
$4\times10^{-4}$ for all components. Our modelling essentially yields the CO column density for each
component, which is proportional to \mdot\,$f_{CO}$, hence the inferred
values of \mdot can be easily scaled for a different value of $f_{CO}$.

\subsection{Model Results} \label{sec-analy-model}

We performed model fits for the observed equivalent widths of the CO R1, R2, R5, R7, P1, and
P3 transitions for the H1, H2, H3 and H5+ components for each epoch, except 1989 Jan 19,
because we 
did not have associated calibration spectra for the latter. The measurement of these widths
from our data is described in Appendix\,\sect{obs-ew}. The equivalent width data for
the H4 component were not modelled since this component is very weak and has large
relative uncertainties. 
For each component, we found the range of parameters that
yield model equivalent widths which are all within $\pm\,Q\sigma$ of the measured
values, where $\sigma$ is the measured uncertainty in the equivalent width values
and $Q$ is a ``tolerance" factor. Each fit was performed with a tolerance of
$Q=1.5$, except for the H1 component from 1988 Nov 30, which could only be fit with
$Q=2$. We stepped each parameter's value in small increments through a large
parameter space defined by reasonable minimum and maximum values, which are listed
in Table \ref{tbl-range}. 
%{\bf The other parameter limits are guided by (BEN:insert text and REFS here!!!)}. 
Once the full parameter space was explored, the
range of acceptable values was determined as the weighted average of all those which
satisfied our fitting criterion as described above, using standard $\chi^2$
weighting.
%({\bf BEN: weighting of what?})

The best fit values of
the 5 model parameters, and the column
density $N$ inferred from these for each high-velocity component, are summarized in Table
\ref{tbl-model}. Although the model
fitting was carried out for each epoch, the differences in the fitted parameters
from epoch to epoch are within the uncertainties of the values derived for each epoch, hence
only average values of the
parameters are provided. 

For the three major kinematic components (H1, H3, H5+) of the HV outflow, the ratio of the
mass-loss to average
radius lies in the range $2.4-4.4$ (in units of $10^{-5}$ \msunyr/$10^{16}$cm), their 
column densities are $\sim (1-4)\times10^{17}$ cm$^{-2}$, and the fractional
shell thickness is about 0.7--0.9 (Table\,\ref{tbl-model}). The temperature at the average
radius ranges from about 115
to 575\,K, and the exponent of the power-law temperature gradient is in the range $-1.5$ to
$-2$.

\subsubsection{Mass-ejection Rates}\label{massloss}
We can derive mass-loss rate values for each of the high-velocity kinematic components, if we
can constrain the average radius of the shells producing these. SW88 found that the HV
features had to be produced in a region with an angular size smaller than $\sim2.5{''}$,
corresponding to $1.4\times10^{16}$ cm at a distance of 390 pc \citep{RH83}. If we assume the
upper limit for the thickness of the shells based on the time-variations, i.e., $10^{16}$ cm,
then, using our model values of the shell thickness to radius ratio, the typical shell radius
is
$<10^{16}$ cm. Hence the radius of the shells appears to be $<10^{16}$ cm,
based
on two independent arguments. The mass-loss rates of the major components (H1, H3, H5+)
thus lie in the range $>(2-4)\times10^{-5}$ \msunyr.

But note that these values of the mass-loss rate are really proxies for the absorption column
density, and do not provide the actual mass-ejection rate per year, $\dot{M}_{HV}$, for the
material
observed in the HV 4.6\,\um~absorption features. The latter can only be calculated by
scaling our model values by the solid angle, $\omega$ (as a fraction of 4$\pi$)
subtended by the absorbing material at the star -- our data do not provide this
information. However, given the highly collimated nature of the HV outflow, it is
likely that $\omega$ is significantly smaller than unity. We can make a rough estimate of
$\omega$ by noting that the values of the (projected) opening angle measured for the
optical outflow (20\arcdeg--30\arcdeg, Setal03), and
that (24\arcdeg) derived from the size of the most distant, high-velocity blue-shifted CO
J=2-1 blob ($5{''}$ at a radius of about $12.5{''}$; Hetal04), are very similar. Hence, we use a
deprojected opening angle of 25\arcdeg\,Sin$i$ (where $i$ is inclination angle of the fast
outflow to the line of sight) for the shells producing
the HV 4.6\,\um~absorption features. We find that $\omega=1.2\times10^{-3}$Sin$i$; taking
$i$=25\arcdeg (mean of the values suggested by Setal03 and Hetal04), we obtain 
$\omega=5.2\times10^{-4}$, and the mass-loss rates for the major HV components, $\dot{M}_{HV}
>(1-2)\times10^{-8}$\msunyr .

In comparison, the mass-loss rate for the HV outflow estimated from the single-dish CO J=3-2
and 2-1 data (KJY97) is larger by more than 2 orders of magnitude. They compute this rate by
deriving the
total mass of the high-velocity outflow from the CO fluxes using a constant density LVG code,
and
then dividing this by an expansion time-scale derived from the outer radius, assumed to
be $5\times10^{16}$ cm, a value supported by the CO J=2-1
position-velocity maps by Hetal04 which show the high-velocity outflow extending to about
10$''$-12.5$''$, or $(5.7-7.3)\times10^{16}$ cm on either side of the center. KJY97 find, for
the high-velocity outflow, \mdot$=7.4\times10^{-6}$\,\msunyr, or $1.85\times10^{-5}$\,\msunyr 
after
scaling to our lower value of the fractional CO-to-H$_2$ abundance ratio. The discrepancy
between the HV mass-loss rates as derived from the mm-wave lines and the 4.6\,\um~lines
implies that only a small fraction of the total HV gas in V Hya is at the high temperatures we
derive (i.e., $>100$\,K).

\subsubsection{Kinetic Temperatures}\label{mod-temp}
The kinetic temperature in the outflows is extremely sensitive to the
ratio of the absorption intensities of high-$J$ to low-$J$ transitions.
%Since we fixed the equivalent width of the H2 component for R5 and R7,
%the derived temperatures are really rough upper limits.  
We note that the H2 component is produced in markedly cooler gas, compared to the
other three components.
We find that, with the exception of the H2 component,
the outflow temperature increases with decreasing expansion velocity of the
outflow.
A plausible explanation for this trend is, assuming the outflows
are gas parcels (``bullets") which are ejected with the same 
velocity from the center, that the slower components are those which have interacted
more extensively with the ambient circumstellar material, and thus have had a larger fraction
of their kinetic energy converted into thermal energy.
The derived temperature dependence $r^{-1.5\pm1.0}$
shows that there is a significant radial temperature gradient in each
component, as one may expect in regions of strongly-shocked gas (e.g., Lee \& Sahai 2003).
%REFER TO HYDRO SIMULATIONS SUPPORTING RADIAL GRADIENT IN SHOCK STRUCTURES!!!

\subsubsection{Line-Widths}\label{mod-lw}
In our model fitting, we have allowed the intrinsic line width, $\Delta V_t$, to act as a free
parameter. We can thus compare the model values of $\Delta V_t$ to the observed FWHM values of
the absorption features as an independent test of our modeling. 
%The turbulent line width data is given as the average and 
%standard deviation of the six FWHM for a given component and epoch. 
We find that the model line width values (6.4, 2.4 and 6.5 \kms~for H1, H2, H3;
Table\,\ref{tbl-model}) agree reasonably well (i.e., within $3\sigma$) with those derived from
fitting the line profiles ($7.9\pm1.2$, $3.8\pm0.8$ and $8.8\pm1.8$ \kms~for H1, H2, H3),
taking into account the intrinsic resolution of the spectra, which is about
2.8\,\kms. For the H5+ component, which is a combination of H5 and H6, the model $\Delta V_t$
is 9.1\,\kms. In comparison, H5 and H6 have $\Delta V_t$ values of $7.5\pm3$ and $7.9\pm1.8$
\kms, which combined together, give $10.9\pm3.5$ \kms. 

Thus both our observed and model values
of the intrinsic line widths are quite large compared to the typical values ($\sim1-2$ \kms)
used for modelling normal AGB CSEs. Low values of $\Delta V_t$ result in the model lines
becoming very optically thick, making their equivalent widths relatively insensitive to
changes in column density or temperature. In this very optically thick regime, a good fit to
the data, which show that the low-$J$ lines have significantly smaller equivalent widths than
the high-$J$ lines, is not possible. Reasonable fits are obtained only when $\Delta V_t\gtrsim
5$ \kms (except for H2). 

The presence of large turbulent velocities is physically intuitive,
since the energetic hydrodynamic interaction of the HV gas with the slowly expanding dense
outflow will produce shock-waves and instabilities which increase the turbulence in both the
HV and the normal outflow. But there is a caveat to this explanation of the large line-widths
-- there may be a significant contribution to the latter due to systematic  velocity gradients
in the absorbing columns for each component. However, both large intrinsic line-widths and
large velocity gradients have the same qualitative effect on (i.e., a reduction in) the
optical depth at any particular velocity, so we do not expect that our model results will be
affected very significantly were we to include a velocity gradient in our model.

\subsubsection{Absorption Feature Overlaps}\label{mod-ovlap}
The different absorption features extracted using our Gaussian fitting overlap in their wings. These
overlaps will result in radiative interaction between the shells producing the absorption for each
of these features. Our model fits the resulting equivalent widths of each of these features
independently, implicitly ignoring this radiative interaction -- such a procedure is acceptable if
the optical depths are small, but if they are $\gtrsim1$, then the uncertainties in the derived
results could be significant. However, radiative interaction between
the different components is not a significant source of uncertainty for our modelling results
because (i) the intrinsic peak
optical depth of each component is modest, and (ii) the overlap occurs at velocities close to or
beyond the half-power point for each Gaussian component, where the optical depths are significantly
lower than those at the peak. For example, for the deepest absorption features (R7, H6 and R5, H3),
the average (over epochs) ratio of the intensity at the center of the line to the continuum is about
$0.6$, 
%even though lines look a bit deeper at their centers in Fig 1, note that each dip has a 
% contribution from a neighbouring line; 0.6 is what we get from results of Gaussian fits
hence the largest
peak optical depth is about $0.5$. The velocity ranges (relative to the systemic velocity) spanned
by the three strongest components H1, H3, and H5+ over their measured widths (FWHM values, corrected
for the instrumental resolution), respectively are $-95$ to $-87.9$, $-85.7$ to $-77.4$, and $-73.0$
to $-62.5$\,\kms, hence
the overlaps of these components occur beyond their half-power range. Since the Gaussian function
representing each component falls rapidly beyond its half-power point, the optical depth of each
component in the overlapping velocity range is quite small. Only the H2 component, which spans the
velocity range from $-87.0$ to $-84.4$\,\kms, has a closer overlap than the others with its
neighbouring component, H3. However, since the H2 component is optically thin (with optical depths
$\lesssim0.15$), it is unlikely to affect the H3 component.

The inferred fractional widths of the shells for the H1, H3 and H5+ components are only modestly
less than unity ($\delta r$ in Table\
\ref{tbl-model}), i.e., the shells are geometrically thick. Our modelling does not determine the
actual average radius for each of the shells, but in order to avoid spatial overlaps between these,
their average radii must be spread over at least a factor of about 5.5. For example, a possible
shell configuration
which avoids spatial overlaps and satisfies the criterion that $\rave<10^{16}$\,cm (see \S\
\ref{massloss}), has
$\rave=1.8\times10^{15}$, $4.4\times10^{15}$ and $9.9\times10^{15}$\,cm for the H5+, H3, and H1 
components\footnote{We have chosen a physically plausible configuration in which the
average shell temperature, $T_{\rave}$, decreases with radius, but this is not required}. We note
however, that since $\delta r$ depends on the
exponent of the model density power-law, $\rho(r)\propto r^{-p}$ ($p=2$ in our modelling), and
steeper exponents will result in smaller values of $\delta r$, the required spread of average radii
could be less than 5.5, if $p>2$.

\section{Discussion}\label{sec-disc}

V Hya is believed to be at a short-lived but critical stage in the evolution of a mass-losing
AGB star into a bipolar PN \citep{Tsu88,Kah96}, and has been dubbed a ``nascent" PPN (Sahai
2007). Based on their imaging survey of young PNs, Sahai \& Trauger (1998) proposed that
collimated fast winds (CFWs) are the primary mechanism for the dramatic change in
circumstellar geometry and kinematics as stars evolve off the AGB. Such jets have been invoked
by Soker (1990) to explain the presence of ansae in PNs. Sahai et al. (2007) carried
out a survey of pre-planetary nebulae,
and found strong similarities in morphologies between PPNs and young PNs, supporting Sahai \&
Trauger's
hypothesis that the CFW's begin operating during the PPN phase or even earlier, during the
late AGB phase. V Hya provides the best and most detailed example of a (carbon-rich) AGB star
in which the CFW's, which are hypothesized to begin the shaping process, are so clearly
manifest.

The near-infrared observations presented here probe the CFW on intermediate scales
($\sim$1000\,AU), beyond the very small scales probed by the optical data of Setal03
($<0.4{''}$ or 150 AU). The interferometric mm-wave CO data mostly probes larger scales,
between a few arcseconds to $\sim15{''}$. These three datasets clearly attest to the episodic
nature of the
V Hydrae HV outflow. Blobby emission is seen in both the CO interferometric maps of the HV
outflow (panels $a$,$c$ of Fig. 2 in Hetal04), and the optical HV emission (Fig. 1 of
Setal03). The multiple kinematic components which we find from our 4.6\micron data directly
show that the HV outflow is discontinuous in velocity space, and, as our modelling results
indicate, the absorbing matter is organised (at least radially) into discrete spatial
structures. We conclude that the HV outflow in V Hya has consisted of discrete blobs (i.e., 
bullets)
throughout its observed history ($\sim$100-250\,yr). These bullet-like ejections are actively 
sculpting the mass-loss envelope of V Hya from the ``inside-out", into a bipolar PPN. V Hya is
not alone amongst AGB and post-AGB objects in which high-velocity outflows appear to be
organised in knots/bullets: other prominent examples are the PPNs, Hen\,3-1475 (e.g.,
Vel{\'a}zquez et al. 2004) and IRAS\,22036+5306 (Sahai et al. 2003b), and the PNs, Hen\,2-90
(Sahai et al. 2000) and MyCn18 (O'Connor et al. 2000).

V Hya is thus a key object in helping us understand how aspherical planetary nebulae are
formed, and should be the focus of new observational and theoretical efforts. For example, a
renewed effort using STIS to follow the evolution of the optically-detected high-velocity
outflow should be made, assuming that the upcoming Hubble servicing mission repairs
STIS\footnote{such a program spanning 3 years was unfortunately cut short due to STIS's
demise}. Near-infrared interferometry at milliarcsec resolution (i.e., few AU at V Hya's
distance) using the VLTI AMBER instrument should be used to probe the central engine which is
producing the high-velocity bullets. 3-D hydrodynamic numerical simulations of such
bullet-like outflows interacting with an AGB mass-loss envelope (e.g., Dennis et al. 2008)
should be carried out to see
if they can reproduce the salient observational features of the high-velocity outflows in V
Hya.

\section{Conclusions}\label{sec-concl}

We have obtained high-resolution spectroscopic data at 4.6\um of the carbon star V Hya, using 
the Fourier Transform Spectrometer at the coud\'e focus of the Kitt Peak 4-m Mayall
telescope. These data, taken over 7 epochs spanning 2 years, cover a large number of P-- and
R--
branch transitions in the fundamental vibration-rotation band of CO and its isotopomers. We
find:

(i) The circumstellar environment of V Hya consists of a
slow-moving normal outflow at about 10\,\kms, and a complex high-velocity 
outflow containing at least six kinematic components with
expansion velocities ranging between 70 and 120 \kms. Weaker absorption can be seen extending
to blue-shifted velocities upto 140\,\kms.

(ii) Physical changes occur in the high-velocity outflow regions on a time-scale as short as
two days, which constrains their extent to be $<10^{16}$ cm.

(iii) The absorption in the high-velocity features is not optically thin, i.e., we cannot assume
that $\tau << 1$. The strongest features have typical optical depths of about 0.5. Using a
non-linear, multi-parameter best-fit model, we can characterize each of the
kinematic components in terms of its measured expansion velocity, the ratio of the mass-loss
rate to the average
shell radius (a proxy for the absorption column density), the fractional shell thickness, a
temperature power-law, and an  
intrinsic line width. 

For the three-most intense kinematic components, the
column density is $\sim (1-4)\times10^{17}$ cm$^{-2}$; the fractional
shell thickness is about 0.7--0.9; the temperature at the average radius ranges from about 115
to
575\,K, and the exponent of the power-law temperature gradient is in the range $-1.5$ to $-2$.

(v) The (projected) opening angle of the outflow (25\arcdeg), does not vary significantly from
small ($0.25{''}$) to large radii ($12.5{''}$), as sampled by the optical and millimeter-wave
observations. Assuming the same value for the high-velocity outflow seen in the 4.6\um data
(and an
inclination angle to the line of sight of 25\arcdeg), we find that the mass-ejection rate of
material seen in the major HV outflow components lies in the range
$\gtrsim(1-2)\times10^{-8}$ \msunyr.

(iv) Both the observed and model values of the intrinsic line-widths are quite large
($6-8$\,\kms~in all except one kinematic component) compared to the typical values ($\sim1$
\kms) appropriate for the circumstellar envelopes of normal AGB stars. The large widths are
probably due to turbulence broadening, and/or systematic velocity gradients, resulting from 
the energetic hydrodynamic interaction of the high-velocity gas
with the slowly expanding dense outflow.

\acknowledgements
We thank an anonymous referee for helpful comments. RS's contribution to the research described in
this publication was
carried out at the Jet Propulsion Laboratory, California Institute of Technology, under a
contract with NASA. RS thanks NASA for financial support via $HST$ awards GO-9632.01 and
GO-9800.01 
administered by STScI, and an LTSA award. BEKS acknowledges support from the Department of
Physics at Occidental College and the Caltech SURF program.

\appendix

\section{Calculating the Model Equivalent Widths}\label{app-ew}

We compute the absorption intensity for each component in a particular
transition $(v=0,J)\rightarrow(v=1,J')$ at a given velocity as
follows.  The absorbing shell is divided into a large number of
very thin radial zones.  The attenuation of incident radiation by each
of these thin shells (at a given velocity) can be written as
$\exp\left[-\Delta \tau(r_i,V)\right]$, where $\Delta \tau(r_i,V)$ is
the radial optical depth due to the $i^{\mathrm th}$ thin radial zone.
The optical depth is proportional to the column density, the $J$-level
population in the vibrational ground state $(v=0)$, the infrared
matrix element for dipole transitions, and the Gaussian velocity
function. 

The column density $N$ of CO in the $i^{\mathrm th}$ thin radial zone is given by:
\begin{equation}
 N_i=\left(\frac{1}{r_{i,-}}-\frac{1}{r_{i,+}}
 \right)\frac{\dot{M} f_{CO}}{4\,\pi\,V_{exp}}, 
\end{equation}
where $r_{i,\pm}$ are the outer and inner limits of the $i^{\mathrm
th}$ thin radial zone and $f_{CO}$ is the number abundance ratio of CO to
H$_2$.  The populations of the different $J$ levels in the ground vibrational state is found,
assuming LTE, using the partition function
\begin{equation}
 n_J = \frac{g_J \exp(-E_J/kT)}{Z}
\end{equation}

where $g_J=2J+1$ is the statistical weight, and 
\begin{equation}
 Z=\sum_{J=0}^{\infty} g_J \exp(-E_J/kT).
\end{equation}

%The IR dipole matrix element has been determined from the matrix
%element
%\begin{equation}
% \left\langle L \left| \frac{dD}{dx}\right| U \right\rangle
%\end{equation}

%where $L$ and $U$ are the lower and upper $J$ states, and
%$\frac{dD}{dx}$ is the change in electric dipole moment of the CO
%molecule with internuclear separation $x$.  

The velocity dependence is described by the Gaussian velocity function
\begin{equation}
 \phi(V)=C\exp{\left[-4\ln{2}\left(\frac{\Delta V}{\Delta V_t} \right)
\right]}
\end{equation}
with $\Delta V=V-V_0$, where $V_0$ is the line center velocity and $C$
is the normalization constant equal to $(\sqrt{4\ln{2}/\pi})/\Delta V_t$, and $\Delta V_t$ is the
intrinsic line width.

The fractional absorption intensity at a given velocity is simply the
product of the attenuation factors due to each thin radial zone
\begin{equation}
 f(V)=\prod_{i=1}^{N}\exp{\left[-\Delta \tau(r_i,V) \right]}
\end{equation}
The equivalent width is then given by 
\begin{equation}
 W_\nu=\frac{\nu}{c}\int{\left[1-f(V)\right]dV}
\end{equation}
where the integration range is chosen to be sufficiently large that
$1-f(V)\approx 0$ near the boundaries of this range.  

\section{Measuring the Observed Equivalent Widths}\label{obs-ew}

We calculate the 
equivalent width of each absorption component by fitting
Gaussians to each of the six HV absorption features. The spectra were first
smoothed
using a three-point (Hanning) filter that averages 50\% of each channel with
25\% of the adjacent lower and higher channels.
Our data reduction package included a non-linear, least squares
Levenberg-Marquardt algorithm which allowed the deconvolution of up to
three Gaussians at a time.  Assuming that there are no HV components
blueward of H1, we make a simultaneous 3-Gaussian fit to the triplet consisting of H1 and
its two redward neighbors, H2 and H3.  The fits for H1 and H2 were then 
subtracted, leaving H3.  In this way we account for
Gaussian superposition on the blue side of H3.  We then repeat the above 
process for the H3--H5 triplet, and finally
fit the H5--H6 components. 
%An example of a fit with all 6 components is shown in Fig.\,\ref{hivfit}. 
A particular fit to all six components is
considered acceptable if minor perturbations in the input parameters
yield the same output fit parameters.  For all epochs, the residual spectral
artifacts following subtraction of all Gaussians were less than 3.5\%
of the continuum level. Uncertainties in the equivalent widths were
estimated by allowing the continuum level to vary by 3.5\%, and
propagating the effect of this variation on the peak and FWHM of each
fitted component.  
 
In some cases, the deconvolution process was not as straightforward as
just described, because the absorption-line features are too closely
spaced for a unique deconvolution.  To address this issue, we assume that the
line centers and FWHMs (in velocity space) of such features will not vary significantly
over different $J$ transitions. For example, since the R5 and R7 features of the H2 components  
could be properly separated, we fixed the centers and FWHM of their Gaussians
to be the same as the lower-$J$ transitions.  Similarly, the lines in the P1 and P3 transitions
of the H4
component could not be uniquely deconvolved, hence their
centers and FWHM were fixed to be those of the corresponding
R-transitions.  

The velocity separation between the H5 and H6
components is not always large enough to uniquely deconvolve these
components.  Therefore, although we fitted Gaussians to each component, we used the sum
of the derived areas under the fitted line profiles to equivalent width, labeled ``H5+'', for
model fitting.  The expansion
velocity of H5+ is taken to be the average of the velocities of H5 and
H6 and the FWHM is the square root of the sum of the squares of the
individual FWHM.

%%%%%%%%%%%%%%%%%%%%%%%%%%%%%%%%%%%%%%%%%%%%%%%%%%%%%%%%%%%%%%%%%%

\clearpage
%\begin{widetext}

\clearpage

\clearpage

\clearpage

\clearpage

\clearpage

\clearpage
\begin{deluxetable}{l c c c c c}
\tablewidth{0pt}
\tablecaption{Parameter Ranges for Equivalent Width Modeling
  \label{tbl-range}} 
\tablehead{
 \colhead{Limit} & \colhead{\mdot/$\rave$} & \colhead{$\delta r$} &
  \colhead{$T_{\rave}$} & \colhead{$\beta$} & \colhead{$\Delta V_t$} \\
 \colhead{} & \colhead{$10^{-5}$ \msunyr/$10^{16}$cm} & \colhead{} & \colhead{K} &
\colhead{} & \colhead{\kms} 
}
\startdata
Minimum & 0.005& $5\times10^{-4}$ & 1    & 0  & 1        \\
Maximum & 4000 & $1$              & 1000 & 10 & 10 \\
\enddata
\end{deluxetable}

\begin{deluxetable}{l c c c c c c c}
\tablewidth{0pt}
\tabletypesize{\footnotesize}
\tablecaption{Model Fitting Results \label{tbl-model}}
\tablehead{
 \colhead{Parameter} & \colhead{H1} & \colhead{H2} & \colhead{H3} &  \colhead{H5+} \\
 \colhead{$V_{exp}$\tablenotemark{a}} & \colhead{91.4} & \colhead{85.8} & \colhead{81.5} &
\colhead{67.8}    
}
\startdata
\mdot/$\rave$      & 2.4\tpm0.5& 0.74\tpm0.16  & 4.4\tpm0.7  & 3.3\tpm0.6  \\
$\delta r$         & 0.72\tpm0.12& 0.25\tpm0.05& 0.90\tpm0.14& 0.72\tpm0.12\\
$T_{\rave}$        & 116\tpm20 & 32\tpm19      & 263\tpm28   & 574\tpm38   \\
$\beta$            & 1.8\tpm0.4& 1.5\tpm0.4    & 1.5\tpm0.4  & 1.4\tpm0.4  \\
$\Delta V_t$       & 6.4\tpm0.7& 2.4\tpm0.5    & 6.5\tpm0.6  & 9.1\tpm0.7  \\
$N$                & 0.9\tpm0.3& 0.1\tpm0.02   & 3.7\tpm0.6  & 2.3\tpm0.2  \\
\hline
%%%%%%%%%%%%%%%%%%%%%%%%%%%%%%%%%%%%%%%%%%%%%%%%%%%%%%%%%%%%%%%%%%
\enddata
\tablenotetext{a}{$V_{exp}$ for each component is determined by Gaussian profile fitting as
described in \S\,Appendix\,\ref{obs-ew}}
\tablecomments{Units are: $V_{exp}$: \kms, \mdot/$\rave$: $10^{-5}$ \msunyr/$10^{16}$
cm,  $T_{\rave}$: K, $\Delta V_t$: \kms, $N$: $10^{17}$ cm$^{-2}$, $\delta r$:
dimensionless}
\end{deluxetable}
%\end{widetext}

\end{document}